# Phosphorene Nanoribbon Based Nano-Electrodes for Explosives Detection: A DFT Study

Rameshwar L. Kumawat[†,], Biswarup Pathak[*,†,#,]

[†]Discipline of Metallurgy Engineering and Materials Science, and [#]Discipline of Chemistry, School of Basic Sciences, Indian Institute of Technology (IIT) Indore, Indore, Madhya Pradesh, 453552, India

*E-mail: biswarup@iiti.ac.in

**Abstract:** We have explored the possibility of using low-dimensional phosphorene based nanoscale device for the detection of nitroaromatic-based explosives. In this work, we have investigated the structural, electronic, adsorption, and quantum transport properties of phosphorene nanoribbon (APNR) in the presence of different nitroaromatic compounds (NACs) using the state-of-the-art first-principle density functional theory (DFT) calculations. Our results reveal that the adsorption of explosive molecules is far from the APNR surface, which neither affects the structure of the explosive molecule nor the APNR surface. However, it changes the electronic energy-gap due to the charge transfer between the APNR and explosive molecule. Furthermore, we have examined the transmission function and the current-voltage ($I-V$) characteristic curves for the APNR+explosive systems with the APNR device as a reference employing the non-equilibrium Green's function (NEGFs) combined with DFT approach. The different current-voltage characteristics (compared to pristine APNR device) of the system in the presence of explosive molecules indicate that such APNR based device can be very much sensitive and selective towards certain explosive molecules. Hence, our study indicates that APNR material may be an attractive nanodevice for the detection of explosives.

## 1. Introduction

The increasing demand for reliable detection of explosive materials has become heightened importance in recent years for various reasons (such as for homeland security, counterterrorism, military monitoring and so on).[1] Therefore, explosive detection is very important to locate landmines as well as improvised explosive devices (IEDs). Furthermore, locating such unexploded ordnance is also a difficult task in homeland security and counterterrorism activities.[2,3] Therefore, it is essential that explosive material based devices need to detect a wide range of explosive materials; such as traditional high explosive materials as well as the chemical found IEDs. Traditional techniques for explosive detection include chemical and electrochemical sensors, luminescence sensors, and spectroscopic techniques to name a few.[1-5,7]

However, the major challenge with the present techniques is in detecting the unexploded ordnance with traditional most explosive vapour materials. Nitroaromatic compounds (NACs) are such explosive based materials, which are mostly used in IEDs.[3,5] Therefore, the detection of vapours coming from the trace level of explosives from the explosive device infers us the possible existence. Hence direct detection of explosive is essenstial because it allows stand-off detection of concealed explosive materials. Also, it is essential to identify the trace level of explosives developed from explosive materials. The current progress in the detection techniques leads to the improvements of traditional methods and outcomes in the fast development of future explosives detection devices. Moreove, developments for nanoscale device, improvements in stand-off distances, field-ruggedization, portability, selectivity as well as sensitivity have been essential to develop and advance device techniques.[1-3] Therefore, the increasing demand for faster, cheaper, efficient, and rapid detection of explosives constantly infers us to search for suitable sensing materials. Nanotechnology and nanosciences have been concreted the role for the development of such advanced explosives detection techniques. Solid-state materials based detection devices play an imperative role in the development and detection mechanism of explosives. Theoretical and experimental studies have been already reported for the detection of explosive materials. These studies have demonstrated that highly explosive materials (for example TNT, picric acid, and so on) can be detected utilizing graphene oxide, graphene nanosheet, carbon nanotube, silicone nanowire, antimonene, graphyne, arsenene, and metal-organic framework to name a few.[3,6,8-12] Two-dimensional (2D) nanomaterials have been considerably attracting attention, and they have been extensively explored for this. Such atomically thin 2D materials are fascinating for various applications as they show interesting chemical and physical properties. Moreover, 2D materials possess huge surface area and improved sensitivity to the environment, which are the key factors for explosive sensing.

Recently synthesised black phosphorus (BP) is the stable and known allotrope of bulk phosphorus. It consists of layers of phosphorene, phosphorene is the name given to a monolayer of BP.[13-17] Semiconducting BP, a monolayer with phosphorus atom has a puckered structural geometry along the armchair direction.[13-15] It possesses high carrier mobility and a finite tunable bandgap, which is very important for such applications. The huge surface area and more inplane conductance make phosphorene principally robust to alter its immediate environment. Moreover, its electrical and optical properties combined with mechanical properties (i.e. robustness and flexibility) turns this material good candidate for fabrication of chemical sensors.[16-19] Theoretical and experimental studies have explored the sensing behavior of the BP. Recent experimental reports have shown that BP nanomaterials can be stable in air and water.[22-25] For example, Pei and co-workers have experimentally demonstrated a controllable method for the fabrication of good quality as well as air-stable BP layers.[25] All these works specify that BP can be a desirable material for explosive detection. Futher, BP based phosphorene nanoribbons[20] are attracting a lot of attention due to their outstanding structural, electronic and transport properties.[21] Hence, inspired by all these reports, we have conceived the idea that phosphorene nanoribbon based nanodevice can be modeled for explosives detection.

Therefore, it is imperative to perceive the adsorption mechanism between the puckered BP nanoribbon and explosives to completely realize the prospects of phosphorene as the explosives detection device. Herein, we have used density-functional-theory (DFT) calculations for the detection of NACs based explosive molecules (picric acid, *m*-dinitrobenzene, hexanitrostilbene, and 2,4,6-trinitrotoluene, as shown in **Figure S1 (Supporting Information)** on the atomically thin phosphorene nanoribbon surface. We have investigated the structural, adsorption, electronic, and transport properties of phosphorene nanoribbon device and phosphorene nanoribbon+explosive molecule systems. Further, we have examined the transmission function and corresponding current-voltage ($I - V$) pattern for different explosive molecules detection on the substrate.

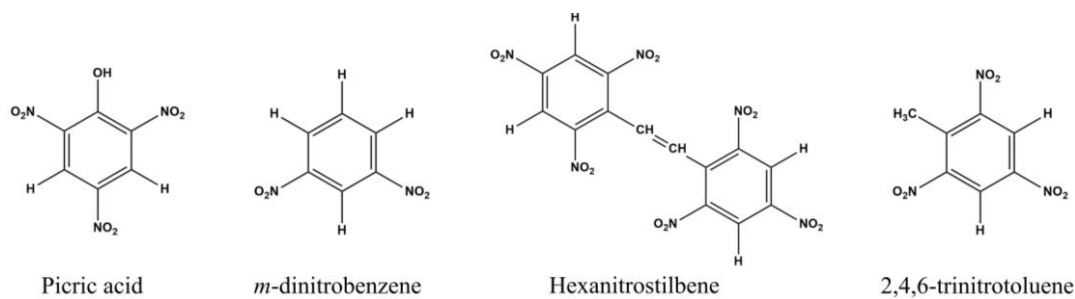

**Figure 1.** Structure and chemical formulas of highly explosive materials.

**2. Model and Computational Methods**

In this work, we have proposed an armchair phosphorene nanoribbon (APNR) based nanoscale sensing device as shown in **Figure 2**. The nanoscale schematic of **Figure 2** shows a sensor concept of the type which inspires the current study, with the APNR serving as sensing of explosive gas molecules. We have taken H-terminated (edges) APNR (**Figure 3**) because H-terminated edges can provide facility to be produced experimentally.[26] To study the structural, adsorption, electronic, and quantum transport properties, we have considered the APNR device and the four explosive vapour molecules for our study: (i) picric acid, (ii) *m*-dinitrobenzene, (iii) hexanitrostilbene, and (iv) 2,4,6-trinitrotoluene.

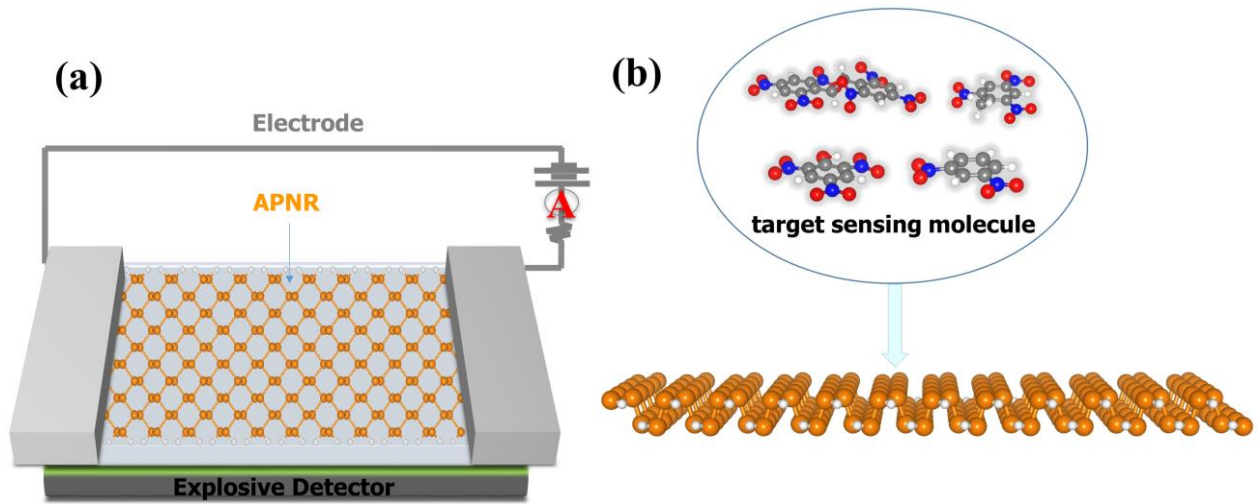

**Figure 2**. Schematic diagrams of (a) phosphorene nanoribbon based nanoscale device, and (b) phosphorene nanoribbon based sensing of explosive gas molecules.

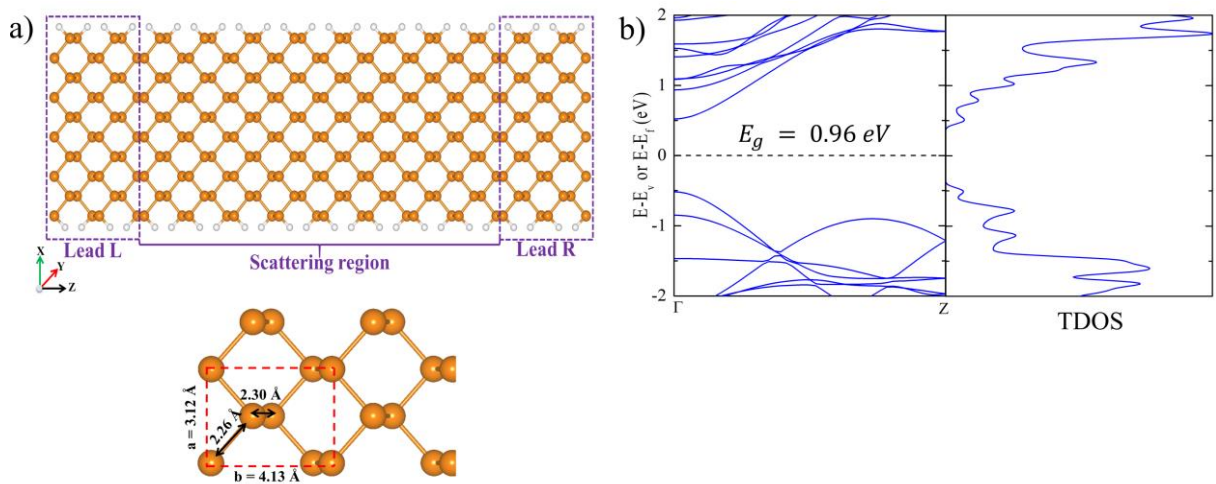

**Figure 3**. (a) Atomic structure proposed APNR nanoscale device illustrating the nano-electrodes (Left (L) lead and right (R) lead) and a central scattering region. (b) The Electronic band structure and the total density of state (TDOS) of the APNR nanoscale device [width (n) = 10]. Here the Fermi-level has been aligned to zero and indicated by the black dashed line.

All the geometries of isolated explosive molecules have been relaxed using the B3LYP/6-311++G** basis sets by Gaussian 09 code.[27-29] Taking the relaxed geometries of explosive molecules, we have adsorbed each explosive molecule on the surface of APNR. For optimization, the hybrid van der Waals (vdW-DF-cx; where cx=BH)[30] forces are used for all calculations as implemented in the SIESTA code.[29] The hybrid vdW-DFs which are developed using 25% Fock exchange with (i) the consistent-exchange vdW-DF-cx (cx is short for consistent exchange)[30] functional and (ii) with vdW-DF2 functional.[31,44] We have also used GGA-PBE level of theory[34] to compare the adsorption energy values and adsorption heights with vdW-DF functional as implemented in the SIESTA.[27,29] Troullier-Martins norm-conserved pseudopotentials are used for core and valence electrons.[33] Double-$\zeta$ polarized basis sets are used for all atoms.[37,32-34] A mesh cut-off value of 300 Ry is used in the calculations. We have been employed a vacuum of ∼32 Å in the $xy$ directions to avoid electrostatic interactions between the repeated images. For the sampling of Brillouin zone integration, we have used Monkhorst k-space grid of $1 \times 1 \times 7$ for the whole system, while $1 \times 1 \times 25$ for the electrode calculations. The electrodes are periodic in the transport direction ($z$-direction). The conjugate gradient (CG) algorithm and tolerance in density matrix difference is $10^{-4}$ and residual forces on the atoms are less than $10^{-2}$ eV/Å have been used for relaxation part.

We have calculated the formation energy of H-functionalized armchair phosphorene nanoribbon (APNR) using the following equation[42]:

$$E_{Formation} = (E_{APNR} - 20 \times E_P - m \times E_H)/m$$

where $E_{APNR}$ is the total energy of the APNR system and $E_P$ and $E_H$ are the per atom energy of phosphorus, and H in the phophorene and $H_2$ structures respectively. Here, m=4, as there are four H binding sites in one primitive unit cell. Our calculated $E_{Formation}$ is -1.08 eV/H for APNR.

The adsorption energies ($E_{ads}$) of the four explosive molecules is calculated using the following equation:

$$E_{ads} = E_{APNR+explosive} - (E_{APNR} + E_{Explosive})$$

where $E_{APNR+explosive}$ represents the total optimized energy of the APNR+explosive device. Here $E_{APNR}$ and $E_{Explosive}$ are the energy of the PNR device and explosive molecule, respectively within the geometry of the APNR+explosive device.

The transmission function and current-voltage $(I - V)$ characteristics are calculated by utilizing the non-equilibrium Green's function (NEGF) approach within the DFT methodology, as implemented in TranSIESTA.[29] The hybrid vdW-DF-cx (cx=BH) sfuncitoanls are used for the transport calculations. Our proposed device has been divided into three parts known as left (L) and right (R) electrodes and a central scattering region. A larger system for transport calculations is considered where each electrode consists of 48 atoms (2 layers) and the scattering region has 192 atoms (8 layers) as shown in **Figure 3**. The transmission function for the APNR system is calculated by considering the Monkhorst k-space grid $1 \times 1 \times 25$ and $1 \times 1 \times 100$ Monkhorst k-space grid in the transport direction. We have found that the magnitude and pattern of the transmission curve do not change much with the higher Monkhorst k-space grid (**Figure S1, Supporting Information**). Therefore, all the bias-dependent transport calculations are done using the $1 \times 1 \times 25$ Monkhorst k-space grid to avoid the computational cost. The electric current through the central scattering region is calculated by exploring the Landauer-Buttiker methodology:

$$I(V_b) = \frac{2e}{h} \int_{\mu_R}^{\mu_L} T(E, V_b) [f(E - \mu_L) - f(E - \mu_R)] dE$$

where $I(V_b)$ represents the current under applied bias voltages, $e$ is the electron charge, $h$ is the Planck constant, $f(E - \mu_{L/R})$ is the Fermi-Dirac distribution functions of the L and R electrodes, $\mu_{L/R}$ ($\mu_{L/R} = E_F \pm V_b/2$) is the chemical potential which can move up and down according to the Fermi energy $E_F$, and $T(E, V_b)$ is the quantum transmission probability of the electrons, which can be given as follows:[19,32,35,38,39]

$$T(E, V_b) = \Gamma_L(E, V_b) \mathcal{G}(E, V_b) \Gamma_R(E, V_b) \mathcal{G}^\dagger(E, V_b)$$

where the coupling-matrices are given as $\Gamma_{L/R} = i[\Sigma_{L/R} - \Sigma_{L/R}^\dagger]$ and the NEGFs for the scattering region given as $\mathcal{G}(E, V_b) = [E \times S_s - H_s[\rho] - \Sigma_L(E, V_b) - \Sigma_R(E, V_b)]^{-1}$, where $S_s$ is the overlap matrix and $H_s$ is the Hamiltonian matrix, $\Sigma_{L/R} = V_{S_{L/R}} g_{L/R} V_{L/R} s$ is the self-interaction energy, $\Sigma_{L/R}$ is a molecule electrode that takes into account from the L/R electrodes onto the central scattering region, $g_{L/R}$ is the surface $L/R$ Green's function and $V_{L/R} s = V_{S_{L/R}}^\dagger$ is the coupling-matrix between L/R electrodes and the scattering region.[19,21,38]

## 3. Results and Discussion

### 3.1 Phosphorene nanoribbon (APNR) nanoscale device

Initially, we have relaxed the unit-cell of the phosphorene (**Figure 3**), and the calculated lattice parameters (a=4.13 and b=3.12 Å) and P-P bond lengths (2.26 and 2.30 Å) are in agreement with the previous reports on phosphorene.[21,35,39] After that, we have modeled the APNR nanoscale device, where the structure of the APNR is armchair along the transport direction ($z$ is the transport direction). Accordingly, the edges of the armchair phosphorene are passivated by hydrogen atoms as shown in **Figure 3**. The calculated band-structure and density of state (DOS) of the APNR setup [width (n) = 10] reveal that it has a bandgap of 0.96 eV, which further agrees well with the previously reported bandgap of ~1.0 eV of the same width.[39,42,43]

Additionally, it shows an underestimation of the bandgap with the use of different vdW-DF forces (**Table S1**). Afterward, the adsorption behaviors of the four explosive molecules [(i) picric acid, (ii) *m*-dinitrobenzene, (iii) hexanitrostilbene, and (iv) 2,4,6-trinitrotoluene] are investigated on the surface of the APNR device. We have examined the possible structural orientations [parallel and perpendicular (**Figure S2-3**)][3,8,11,12] of all four explosive molecules concerning the APNR surface and the most favorable APNR+explosive structures (**Figure 4** and **Figure S2**) are considered for further study. The computed relative energy values are tabulated in **Table S2** which shows that parallel adsorbtion is the most favorable adsorption configurations for all the explosive molecules. This could be because of the high surface area of target molecules which consists nitro (-$NO_2$) groups and benzene rings which stabilize the molecule towards parallel adsorption.

## 3.2 Adsorption of explosive molecules on the APNR device

The interactive nature of the explosive molecules on the APNR surface is examined based on adsorption height, adsorption energy, and electronic structure. The adsorption energies and adsorption heights, calculated with the PBE and hybrid vdW-DF (van der Waals corrections). It has been found that vdW interactions play a crucial role in such studies due to the presence of long-range interactions.[3,30,31,33] Therefore, we have considered the vdW-DF approach for the correction of vdW interactions to the adsorption energies and adsorption heights. The most favorable adsorption configuration (APNR+explosive) for the adsorbed explosives on the APNR are shown in **Figure 4**. The adsorption height ($h$) is measured from the phosphorus (P) atom of the APNR sheet to the nearest atom of the adsorbed explosive molecule, i.e. the shortest distance between the APNR sheet and explosive molecule. For example, in the case of APNR+picric acid, the adsorption height is the shortest distance between P (of APNR surface) and O (oxygen) atom (of picric acid), as presented in **Figure 4**. Our vdW-DF results (**Table 1**) show that the adsorption heights are 2.99, 3.02, 2.76, and 2.75 Å for APNR+picric acid, APNR+*m*-dinitrobenzene, APNR+hexanitrostilbene, and APNR+2,4,6-trinitrotoluene, respectively. One can see that our vdW-DF calculated adsorption energies are significantly higher compared to the PBE level of theory as vdW-DF overestimates the adsorption energy. However, the order of adsorption energy is the same for all four systems. The *m*-dinitrobenzene has the lowest adsorption energy on the APNR surface while hexanitrostilbene has the largest adsorption energy amongst all four molecules. The adsorption energies show that the interactions of hexanitrostilbene are relatively stronger compared to other molecules. The adsorption energy of hexanitrostilbene on to the APNR is significant due to the presence of two trinitrobenzene molecules. The calculated vdW corrected adsorption energy values are ≤ -3 eV for all the explosive molecules except for hexanitrostilbene. This is because, in the case of hexanitrostilbene, six nitro (-$NO_2$) groups interacting with the APNR surface. Therefore, the adsorption energy per nitro group is more or less similar (~ -3 eV) for all the NAC based explosive molecules though the adsorption heights are more than 2.75 Å. Our PBE (without vdW) calculated values are significantly lower compared to the vdW corrected adsorption values. Interestingly, several reports have shown that SIESTA overestimate the energy.[40,41,21] The Inclusion of vdW-DFs gives a significantly better description of non-covalent forces. And the adsorption species considered in our study contain a significant amount of π-electron clouds, especially located over phenyl rings, there is a considerable amount of weak interactions between them and the phosphorene surface, which is also evident from the bond lengths ranging from 2.75 Å to 3.20 Å. Therefore, consideration of vdW interaction is required for an accurate description of the bonding scenario which is responsible for the observed variation in the binding energies between the non-vdW (i.e. GGA-PBE) and vdW (i.e. vdW-DF-xc; xc=BH) calculations. This notion is further supported by the less difference in adsorption energies while the molecules are oriented along the perpendicular direction as the π-electron clouds are now located relatively far from the phosphorene surface compared to the parallel orientation, which brings the interaction to have a less physisorption nature and predominantly arising from the nitro (-$NO_2$) groups of the adsorbing molecules and phosphorene.

**Table 1.** Adsorption heights ($h$, in Å), and adsorption energies ($E_{ads}$, in eV) of the explosive molecules while adsorbed on the APNR surface.

| Explosive Molecules | Parallel Adsorption | | | | Perpendicular Adsorption | | | |
|---|---|---|---|---|---|---|---|---|
| | PBE | | vdW-DF | | PBE | | vdW-DF | |
| | $E_{ads}$ (eV) | $h$(Å) | $E_{ads}$ (eV) | $h$(Å) | $E_{ads}$ (eV) | $h$(Å) | $E_{ads}$ (eV) | $h$(Å) |
| picric acid | -1.05 | 3.21 | -2.78 | 2.99 | -0.32 | 3.01 | -0.74 | 2.91 |
| *m*-dinitrobenzene | -0.96 | 3.18 | -2.40 | 3.02 | -0.41 | 2.93 | -0.83 | 2.89 |
| hexanitrostilbene | -2.02 | 2.83 | -5.10 | 2.76 | -0.87 | 2.90 | -2.11 | 2.76 |
| 2,4,6-trinitrotoluene | -1.29 | 2.85 | -3.10 | 2.75 | -0.42 | 3.06 | -0.97 | 2.94 |

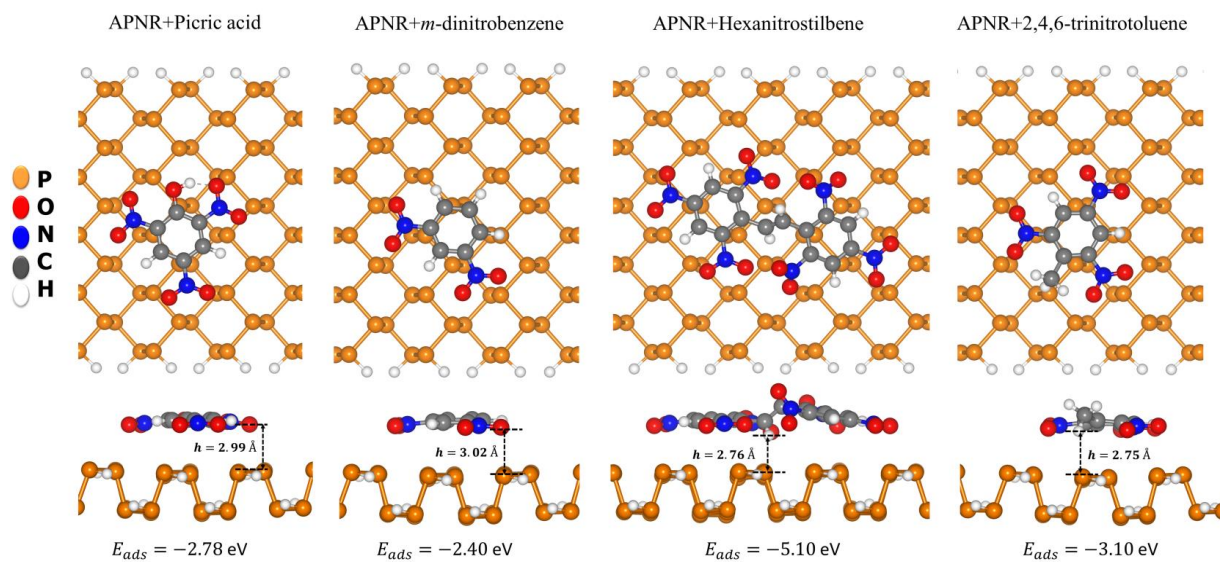

**Figure 4.** Schematic representation of the minimum energy structure (top and side views) of the APNR+explosive systems. Computed adsorption heights ($h$, in Å), and adsorption energies ($E_{ads}$, in eV,) of the APNR+explosive systems are shown.

The computed total/partial density of states (TDOS/PDOS) plots of the APNR device and APNR+explosive molecules are shown in **Figure 5**. **Figure 5** shows that the overlap between NAC explosive molecule and APNR device is less for all the systems, which further indicates that they are interacting weakly. However, the adsorption of explosive molecules changes the Fermi position. Interestingly, we find that the APNR device becomes metallic in the presence of picric acid and hexanitrostilbene. Our PDOS plots show that the main atomic contributions at the Fermi level coming from the phosphorus, not from the explosive molecules. This could be due to the charge transfer from APNR to the explosive molecule, which changes the APNR device to a p-type system. Therefore, the Fermi level shifts towards the valance band for all the cases. Such changes in the electronic properties suggest that the explosive molecules are coupling well with the APNR device and can be utilized for explosive molecule detection. Furthermore, we have investigated the change in transmission function for detecting the explosive molecules from the transmission as well as current-voltage ($I-V$) characteristics.

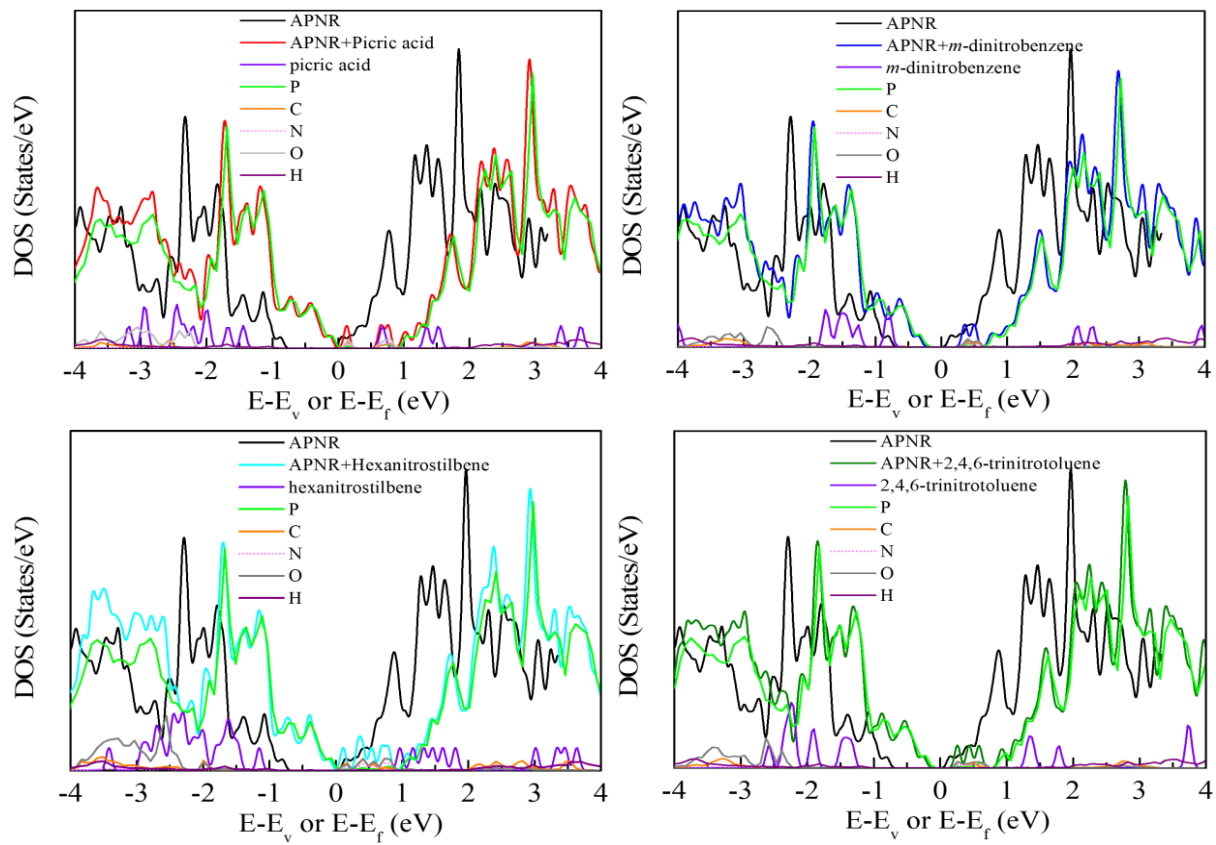

**Figure 5.** PDOS analysis for APNR device and APNR+explosive systems (APNR+picric acid, APNR+$m$-dinitrobenzene, APNR+hexanitrostilbene, and APNR+2,4,6-trinitrotoluene; respectively). PDOS for the isolated molecules and corresponding individual atoms are presented. The Fermi-level has been aligned to zero.

## 3.3 Transmission Function

To understand the sensitivity of the APNR device as an explosive detector, we have examined the transmission function and current-voltage ($I-V$) characteristics for the APNR device in the presence of explosive molecules. **Figure S2** shows the modeled device, which we have used for the transport calculation where the shaded area denotes left (L) and right (R) electrode, and a central scattering region. The central scattering region is the region where the explosive molecules are adsorbed. It has been reported that phosphorene has anisotropic transport behavior and the reported current along the armchair direction is always higher than along the zigzag direction.[36] As adsorption of such explosive molecules leads to charge transfer between the APNR device and explosive molecule, therefore it may enhance/decrease the electrons density on the APNR device compared to that in the pristine APNR device. The transmission function can be a fundamental property for the detection of explosive molecules based on the change in total electron density on the APNR surface. The computed zero-bias transmission function for the pristine APNR and the APNR+explosive molecules shown in **Figure 6**. Figure **6** shows the transmission function around the Fermi energy. The transmission function of the APNR device has step-like behaviour on both sides of the energy region which is very much in aggrement with previously reported work.[45] We have found that there are some distinct molecular states appear in the -0.8 and -1.5 eV energy range, which can be characteristics of the occupied molecular orbital peaks for the different explosives. Similarly, in the energy range between 0.5 and 1.5 eV, we have found very distinct peaks, which can be recognized to the unoccupied molecular orbital peaks. The corresponding density of states (**Figure 6b**) further confirms such distinction.

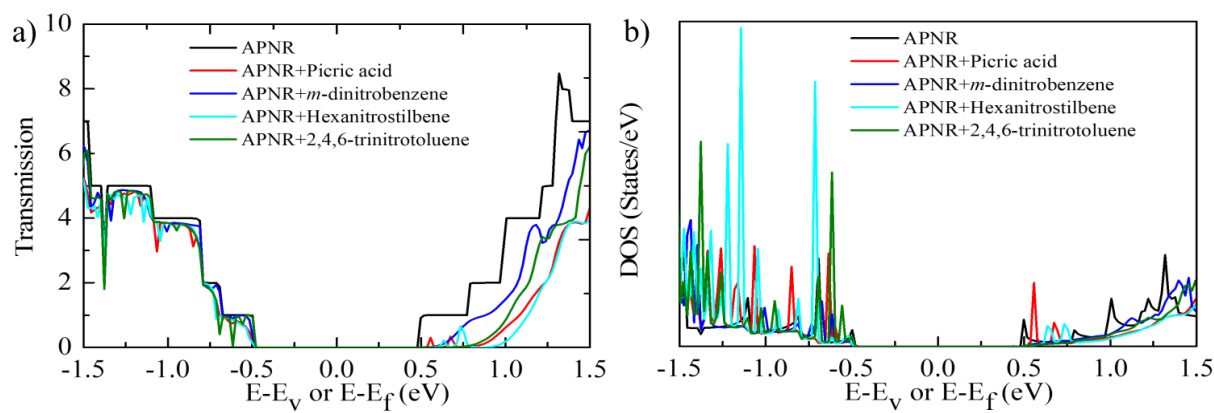

**Figure 6.** (a) The zero-bias transmission function for APNR device and APNR+explosive systems as a function of energy. (b) The corresponding zero-bias density of state (DOS) plotted for APNR and APNR+explosive systems. the Fermi-leve is aligned to zero.

We find that the transmission is lower for the APNR+explosive systems compared to the APNR device. This indicates that the adsorption of explosive molecules leads to back-scattering, which hinders the existing conduction-channels. The change in transmission function is high for hexanitrostilbene, which could be due to the strong adsorption energy. Furthermore, we have investigated the transmission function under an applied bias voltage (V). The computed bias-dependent transmission spectra for all four systems are shown in **Figure S4**. Bias-dependent transmittance unveils that on increasing the bias voltage, the shift in the molecular states can be observed around the Fermi energy. Therefore, the applied bias voltage shows a significant role in the transmission function. As we have discussed above that the adsorption of explosive molecules changes the energy-gap of the APNR device, which results in the change in the transmittance as a function of applied bias voltages.

### 3.4 Current-Voltage ($I-V$) Characteristics

We have calculated the current-voltage (*I-V*) characteristics for the detection of explosive molecules. **Figure 7** presents the $I-V$ curves for all the systems. We find that initially the current (I) remains zero up to an applied bias of ~1 V. This is because APNR is a semiconducting electrode system with a bandgap of 0.96 eV. Therefore, we have to apply a minimum bias of around 0.96 V for the flow of current. As we increase the bias voltage (>

0.96 V), the current value increases. Such $I-V$ patterns can be explained from the transmission peaks under the applied bias window (**Figure S4**). The $I-V$ pattern curve shows that the current value increases continually as we increase the applied bias voltage (upto 2 V). The current signal increases significantly for the APNR+picric acid and APNR+hexanitrostilbene systems compared to other systems.

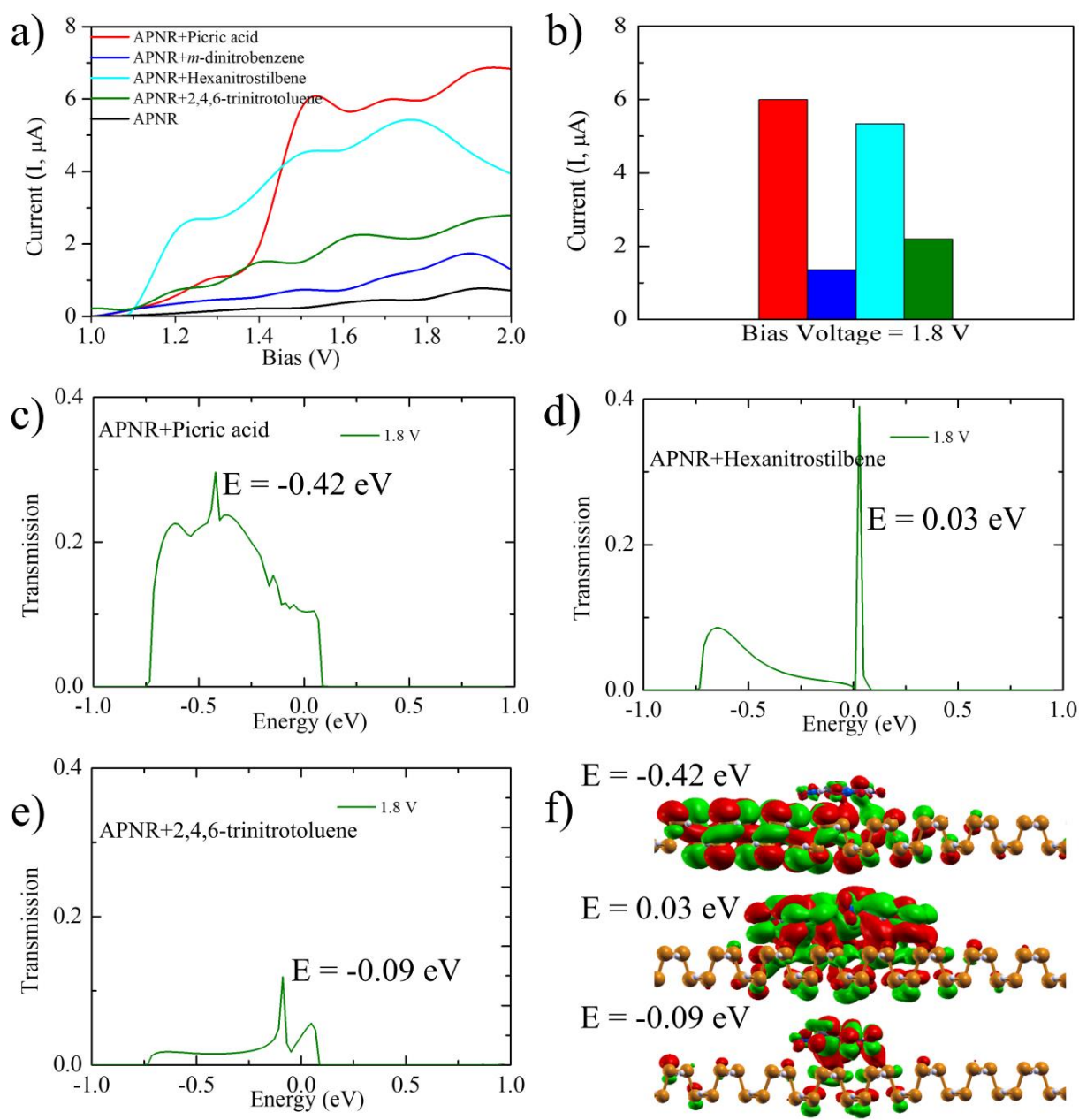

**Figure 7.** (a) $I-V$ curves of APNR and APNR+explosive systems. (b) $I-V$ values for APNR+explosive systems at 1.8 V bias. (c) Transmission function at 1.8 V bias for the APNR+picric acid and MOs responsible for maximum transmission (eigenchannels at -0.42

eV). (d) Transmission function at 1.8 V bias for the APNR+hexanitrostilbene and MOs responsible for maximum transmission (eigenchannels at 0.03 eV). (e) Transmission function at 1.8 V bias for the APNR+2,4,6-trinitrotoluene and MOs responsible for maximum transmission (eigenchannels at -0.09 eV). (f) MOs (isosurface value: 0.05 e/Å$^3$) responsible for the transmission peaks.

Moreover, all systems show different current patterns which could be due to the different structural and electronic behavior of the different explosive molecules. The current scale also indicates a remarkable enhancement in the current value (up to 7 μA) for picric acid. Thus, there is an increase in current signals on the adsorption of all the four explosive molecules. In the 1.2 to 1.4 V bias window, the distinction between all four explosive molecules is not possible though hexanitrostilbene can be distinguished from the other three molecules. However, one can see that the detection of all four explosive molecules is possible in the 1.5-2.0 V bias region. We have also noticed the current signals at 1.8 V (**Figure 7b**) in the following order: $I_{APNR+picric\ acid} > I_{APNR+hexanitrostilbene} > I_{APNR+2,4,6-trinitrotoluene} > I_{APNR+m-dinitrobenzene}$. The current values are high in the presence of picric acid and hexanitrostilbene, whereas low in the presence of 2,4,6-trinitrotoluene and *m*-dinitrobenzene. Hence, our device is sensitive and selective towards picric acid, hexanitrostilbene, and 2,4,6-trinitrotoluene compared to *m*-dinitrobenzene. For the better understanding of transmission and current signals in these three different systems, we have plotted the eigenchannels (**Figure 7c-f**) at 1.8 V bias voltage at -0.42, 0.03, and -0.09 eV. The APNR+picric acid and APNR+hexanitrostilbene systems (**Figure 7**) show a significantly strong overlap between orbitals compared to APNR+2,4,6-trinitrotoluene. This is because of the significant amount of π-electron clouds specially located over phenyl rings and –NO$_2$ groups of these systems which are coupling significantly with the APNR surface. Thus, picric acid, hexanitrostilbene, and 2,4,6-trinitrotoluene can be easily distinguished from *m*-dinitrobenzene due to their strong current signals. In the case of *m*-dinitrobenzene, which possesses eigenstates further away from the APNR electrode's Fermi energy, exhibit different characteristic current magnitude, showing rather a little overlap in the low bias regime. However, at 1.8 V we can clearly detect all four explosive molecules due to their different current values. The presence and absence of these explosives, which is found by different current values, can be considered as ON/OFF state for the APNR based setup. The significant difference in the current signals on account of explosives molecules adsorption is the root of the APNR based detecting setup with enhanced sensitivity.

## 4. Conclusions

In summary, we have investigated the applicability of the phosphorene nanoribbon as a setup for explosive detection. We have systematically studied the structural, adsorption, electronic, and transport properties of the phosphorene nanoribbon device in the presence of four nitroaromatic based explosive molecules such as picric acid, *m*-dinitrobenzene, hexanitrostilbene, and 2,4,6-trinitrotoluene. Investigating a sensing nanoscale device with electrical detection of different explosives, we have investigated the transmission function and the $I-V$ properties for the APNR+explosive systems with the APNR device as a reference. The bias-dependent transmission results suggest that the transmittance has been dominated under applied bias voltages for the APNR+explosive systems. The calculated $I-V$ pattern indicates that the current value increases as we increase the applied bias voltage. The current signal increases significantly for the APNR+picric acid and APNR+hexanitrostilbene systems compared to the other systems. Furthermore, the detection of all four explosive molecules is possible in the 1.5-2.0 V bias region. In particular, picric acid and hexanitrostilbene offer higher current signals, while 2,4,6-trinitrotoluene and *m*-dinitrobenzene offer lesser current. Thus, picric acid, hexanitrostilbene, and 2,4,6- trinitrotoluene is very much sensitive and selective towards the APNR device compared to *m*-dinitrobenzene. Such sensitivity and selectivity of these systems further verified through eigen channel analysis. Therefore, the presence and absence of these explosive molecules can be regarded as ON and OFF states for their detection mechanism. Hence, from these findings, we feel that such sensitivity and selectivity of the APNR device towards selective explosive molecules may make phosphorene an attractive material for explosive detection.

## 5. Supporting Information:

Bandgap with different vdW-DF forces, Atomic structures of proposed APNR+explosive molecule systems, and bias-dependent transmission function for APNR+explosive systems.

## 6. Conflicts of interest

There are no conflicts of interest to declare.

## 7. Acknowledgments

We thank IIT Indore for the lab and computing facilities. This work is supported by DST-SERB, (Project Number: EMR/2015/002057) New Delhi and CSIR [Grant number: 01(2886)/17/EMR (II)]. R.L.K. thanks IIT Indore for research fellowships. We would like to thank Dr. Vivekanand Shukla for fruitful discussion throughout this work.

## 8. ORCID

Rameshwar L. Kumawat: 0000-0002-2210-3428

Biswarup Pathak: 0000-0002-9972-9947

**Table of Content (TOC)**

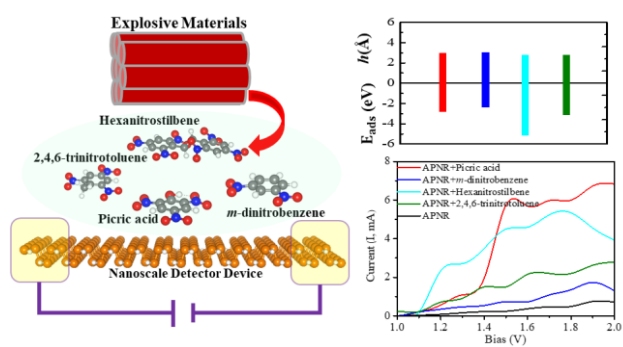